# Electrical Transport Model of Silicene

# as a Channel of Field Effect Transistor

Hatef Sadeghi

Quantum Technology Center, Physics Department, Lancaster University, LA14YB Lancaster, UK

h.sadeghi@lancs.ac.uk

*Abstract*— **The analytical electrical transport model of the Silicene, a single layer of $sp^3$ bonded silicon atoms in the honeycomb lattice structure as a channel in the field effect transistor configuration is presented in this paper. Although the carrier concentration of the Silicene shows similar behavior to Graphene, there are some differences in the conductance behavior. Presented model shows increment in the total carrier and the conductance with the gate voltage as expected for conventional semiconductors which affected by the temperature only in the neutrality point. The minimum conductance is increased by the temperature whereas it remains stable in the degenerate regime. Presented analytical model is in good agreement with the numerical conductance calculation based on the implementation of the non-equilibrium Green's function method coupled to the density functional theory.**

*Index Terms*— Silicene, Electronic Properties, Analytical Model, Carrier Concentration, Conductance

## I. INTRODUCTION

After discovery of Graphene, two dimensional planar sheet of $sp^2$ bonded carbon atoms in the honeycomb lattice, which shows fantastic electrical properties with zero band gap in the Dirac point [1,2], attempts to find one atomic thick structure of the conventional materials such as silicon or germanium raised considerable scientific interest [3-10]. One atomic thick crystalline form of the silicon atoms with $sp^3$ bond arranged in a honeycomb lattice structure so called Silicene (fig. 1*a*), has been experimentally observed by Nakano, et al for the first time [11]. First principles calculation based on density functional theory (DFT) showed that despite carbon atoms that make two demensional planar sheet, Silicene makes stable bounds in the semi two demential (semi-2D) hexagon form [12]. Recently, it has been demonstrated compelling evidence, for the synthesis of the epitaxial Silicene sheets and nanoribbons on the silver (111) [13-20], the Iridium (111) [21] and the zirconium diboride [22] substrates. In addition, a possible superconductivity has been recently observed in the Silicene grown on Ag (111) surface, at 35-40K, which is higher than all other single-element superconductors discovered so far [23]. Similarity of the Silicene and Graphene is due to the same group (IV) that both belong. Larger ionic radius in the Silicene promotes $sp^3$ hybridization [24] implies a small buckling to make Silicene structure more stable whereas $sp^2$ hybridization in the Graphene make it stable in the planar form [25]. This buckling has been reported in the range of 0.36Å to 0.75Å [26-28] by DFT calculation which predicts the band gap up to  800meV [26] too. The gap depends on the external perpendicular electric field  and the amount of the buckling [26] which could be modified by the epitaxial strain [22]. However, this buckling was observed 0.75Å in Silicene grown on Ag [13].

A common problem with the single layer Graphene is its zero gap nature which does not allow proper cut-off in the Graphene based FETs. This issue is not pronounced for the Silicene, since it behaves as a semimetal with variable band gap tuned by the external electric field [29-31], whereas it is transformed to metal for the strain higher than 7.5% [32,33]. However, the electronic properties of the Silicene nanoribbons and Silicene sheets were found to resemble those of the Graphene [34].

More compatibility with existing semiconductor techniques is more pronounced advantage of this material. Recently, Silicene nanoribbon also has been studied [17,18,35-43] which exhibits another advantage of Silicene comparing with the Graphene: the Silicene nanoribbons edges do not reveal oxygen reactivity [44,45]. In addition, $sp^2$-like hybridization of the silicon valence orbitals could be obtained in the Silicene nanoribbons [46]. Calculations predict direct and indirect gap for the zigzag and the armchair edge shape Silicene nanoribbon, respectively [47]. The electronic and magnetic properties of the hydrogen terminated zigzag Silicene nanoribbon decorated with a single carbon chain at the different positions has been also studied which may lead to some applications in switches and sensors [48]. Unlike graphite, which consists of weakly held stacks of the Graphene layers through dispersion forces, the interlayer coupling in Silicenes found to be very strong [49]. Functionalized electronic properties of the Silicene has been studied by Gao, et al. showed that Fluorine and Iodine doped Silicene could open a gap of 1.469 and 1.194 eV respectively which are possible candidates as a channel materials for FETs [50,51]. Also Silicene is stable in a wide doping range [52].

Hydrogenation process could affect the behavior of the Silicene change from metallic to magnetic semiconducting and then to nonmagnetic semiconducting [53-59]. Half-hydrogenation of the Silicene breaks its extended π bonding network, exhibiting ferromagnetic semiconducting behavior with a band gap of 0.95 eV [60]. Moreover, recent calculation shows that Silicene may be potentially used in the spintronic [61,62], nanomagnetic materials [63,64], FET and hydrogen storage devices [49].

However, in the body of the literature, the electronic properties of the Silicene have not been explored yet by systematic analytical theoretical studies. Here the analytical model of the electrical transport including the total carriers, and the conductance of the Silicene as a channel between drain and source (fig. 1*b*) are presented in the different biased and gate voltages as well as temperature to provide better understanding of the behavior of this material as a channel of the FET structure. In addition, its electrical properties have been compared with the Graphene which behaves differently in some aspects. Due to the lack of the experimental studies demonstrating the conductance behavior of the Silicene based FET, the presented result is compared with first-principles quantum transport simulations using computational implementations of the non-equilibrium Green's function method (NEGF) coupled to the density functional theory (DFT) implemented aiding the SIESTA [65] tool. Presented models indicate useful information about the electrical properties of the Silicene in very straight forward form and it can be used to model the Silicene based transistors for the integrated circuits.

## II.  BAND STRUCTURE

Starting point to investigate the electrical properties of a material is looking at its band structure. The spectrum of full tight-binding Hamiltonian of Silicene in the presence of the perpendicular external electric field has been obtained in [66] as shown in equation (1). The differences between electric potential of top $(v_{g1})$ and bottom gates $(v_{g2})$ induce the electric field which breaks the symmetry between A and B sub lattices and hence opening the gap $\Delta = q(v_{g1} - v_{g2})/2$.

$$E = \pm \sqrt{\Delta^2 + \left(\hbar v_F k\right)^2} \tag{1}$$

where $k = \sqrt{k_x^2 + k_y^2}$ is the $k$ vector, $\hbar = 1.054 \times 10^{-34}\ js$ is the reduced Planck's constant and $v_F = 1.3 \times 10^6\ ms^{-1}$ [13,19] is Fermi velocity in the Silicene. Figure 2 shows the band structure of the Silicene in the different gate voltages $(v_g = v_{g1} - v_{g2})$ plotted based on equation (1). As it is clear, the band gap increases with the gate voltage. It is interesting to note that opening a gap in the presence of the electric field is main differences between the Graphene and Silicene which is the consequence of the non-planar structure of the Silicene despite Graphene. In the absence of the gate voltage, Silicene behaves like Graphene [67] and its band structure turns to the Graphene band structure ($E = \hbar v_F |k|$). However it is apparent that the Fermi velocity in Graphene and Silicene are different and even in the absence of the external electric field, they would behave slightly different.

## III.  CARRIER CONCENTRATION IN SILICENE

The carrier concentration in a band is achieved by integrating the Fermi - Dirac distribution function over energy band [68] as $n = \int D(E) f_F(E) dE$ where $D(E)$ and $f_F(E) = (1 + exp(\frac{E-E_F}{k_B T}))^{-1}$ are the available energy states (Density of States) and Fermi-Dirac distribution function, respectively. Density of states could be calculated by summation over the $k$ space with respect to the $E - k$ relation as $D(E) = \sum_k \delta(E - E(k))$. By changing the summation into an integral, the density of states of the Silicene could be written as:

$$D(E) = \frac{WL}{4\pi^2} \int_{-\pi}^{+\pi} d\theta \int_0^\infty k dk \delta(E - E(k)) \qquad (2)$$

where the momentum ($k$) can be obtained from $E - k$ relation of Silicene (eq. 1) as $k = \sqrt{(E^2 - \Delta^2)/\hbar^2 v_F^2}$ which induces $k dk = |E| dE / \hbar^2 v_F^2$. Substituting $k dk$ into the equation (3) and including the spin and the valley degeneracy indicates Silicene Density of States as:

$$D(E) = \frac{2WL|E|}{\pi \hbar^2 v_F^2} \qquad (3)$$

Density of states in cooperation with the Fermi Dirac distribution function indicates the total electron concentration ($n = \int_0^\infty dE D(E)\Theta(|E| - \Delta)f(E)$) of the Silicene as:

$$n = \int\limits_0^{+\infty} dE \left( \frac{2WL|E|}{\pi \hbar^2 v_F^2} \right) \Theta\left(|E| - \Delta\right) \frac{1}{1 + e^{\frac{E - E_F}{k_B T}}} \tag{4}$$

where $\Theta(|E| - \Delta) = \begin{cases} 0, & 0 < |E| < \Delta \\ 1, & other \end{cases}$. By changing the variables as $x = (E - \Delta)/k_B T$

and $\eta = (E_F - \Delta)/k_B T$, the equation (4) could be written as:

$$n = \frac{2WL}{\pi \hbar^2 v_F^2} \left( k_B T \int\limits_0^{+\infty} dx \frac{x k_B T + \Delta}{1 + e^{x - \eta}} \right) \tag{5}$$

The first and the second parts of the carrier concentration model (eq. 5) looks like the complete Fermi–Dirac integral ($\Im_j(x) = \Gamma(j + 1)^{-1} \int_0^\infty \frac{t^j \, dt}{exp(t - x) + 1}$) with the zero and second order ($j$), respectively where $\Gamma$ is the gamma function. Substituting the Fermi Dirac integral into the equation (5), the total electron concentration of the Silicene could be obtained as the equation (6).

$$n = \frac{2WL}{\pi \hbar^2 v_F^2} \left( \Delta \Gamma(1) k_B T \Im_0(\eta) + \Gamma(2) \left( k_B T \right)^2 \Im_1(\eta) \right) \tag{6}$$

Figure 3 shows the electron concentration of the Silicene in the different gate voltages and temperature for a sample with the width and the length of 1μm. The electron concentration increases with the gate voltage as expected in the conventional semiconductors and the rate of the carrier concentration increment is higher for higher gate voltages. However, increasing the temperature dramatically increases the electron concentration in the neutrality point despite higher gate voltage where it remains stable.

## IV.  CONDUCTANCE OF SILICENE

For the Silicene where the electron transport is assumed to be ballistic, the conductance could be calculated using the Landauer's formula in the Ballistic regime [68] $G = (2q^2/h) \int_{-\infty}^{+\infty} dE \, M(E)(-\partial f_F(E)/\partial E)$ where $q$ is the electron charge, $h$ is the Planck's constant. The number of the modes ($M(E)$) that are above the cut-off at the energy $E$ in the transmission channel could be obtained as:

$$M(E) = \sum_k \delta(E - E(k)) \frac{\pi \hbar}{L} \left| v_x(k) \right| \tag{7}$$

where $v_x(k) = dE/\hbar dk_x$ is the carriers velocity in the $x$ direction (fig. 1a) which could be calculated for the Silicene as:

$$v_x(k) = \frac{2\hbar v_F^2 k_x}{\sqrt{\Delta^2 + \left(\hbar v_F k\right)^2}} \tag{8}$$

Substituting the carrier's velocity into the equation (7) and changing the summation to integral by employing $k_x = k\cos(\theta)$ where $\theta$ is the angle between $x$ and $y$ directions in 2D channel and including the spin degeneracy, the total number of the modes in the Silicene channel could be written as:

$$M(E) = \frac{WL\pi\hbar}{4\pi^2} \int_{-\pi}^{+\pi} \left| \cos(\theta) \right| d\theta \int_0^\infty k \, dk \, \delta(E - E(k)) \left| \frac{2\hbar v_F^2 k_x}{\sqrt{\Delta^2 + \left(\hbar v_F k\right)^2}} \right| \tag{9}$$

By justification based on the $E - k$ relation of the Silicene; the equation (9) could be readily written as the equation (10) which indicates the analytical description of the number of the modes in the Silicene 2D channel.

$$M(E) = \frac{W \left| \sqrt{E^2 - \Delta^2} \right|}{\pi \hbar v_F} \qquad (10)$$

In the absence of external electric field ($\Delta = 0$), the number of the modes model would be similar to the number of the modes model in the Graphene as the band structure of the Silicene (eq. 1) would change to graphene like linear energy dispersion when $\Delta = 0$. However, its value would be different in the Silicene due to the different Fermi velocity of its carriers. Therefore, in the absence of the external electric field the number of the modes in Silicene could be obtained as:

$$M(E) = \frac{W \left| E \right|}{\pi \hbar v_F} \qquad (11)$$

The number of the modes in combination with the Landauer's formula indicates the conductance of the Silicene in the absence of the external perpendicular electric field as:

$$G = -\frac{2q^2}{h} \frac{W}{\pi \hbar v_F} \left( \int_0^{+\infty} dE \left| E \right| \frac{\partial}{\partial E} \left( \frac{1}{1 + e^{\frac{E + E_F}{k_B T}}} \right) - \int_0^{+\infty} dE \left| E \right| \frac{\partial}{\partial E} \left( \frac{1}{1 + e^{\frac{E - E_F}{k_B T}}} \right) \right) \qquad (12)$$

Employing partial integration methods, changing variables and some simplifications, the equation (12) could be written as:

$$G = \frac{Wq^2}{\pi^2 \hbar^2 v_F} (k_B T) \Big( \Im_1(-\eta) + \Im_1(\eta) \Big) \qquad (13)$$

The equation (13) describes the zero volt gating conductance of the Silicene. Since the conductance in the Silicene has not been reported experimentally yet, we perform the numerical calculation of the conductivity in the Silicene by the aid of the numerical implementation of non-equilibrium Green's function method coupled to density functional theory to verify presented analytical model. Then the result was compared with the analytical model as shown in the figure 4. The conductivity of the Silicene under different bias voltages obtained from equation (13) illustrated by solid line in the figure 4 as well as the numerical calculation shown by dots. The graph shows good agreement between two methods.

Now let us to consider the conductance in the presence of external perpendicular electric field. The number of the modes, therefore, should be obtained from equation (10). By employing the Taylor series, in the presence of the external electric field ($\Delta \neq 0$), $\left| \sqrt{E^2 - \Delta^2} \right|$ could be extended as: $\left| \frac{E^2}{2\Delta} - \Delta \right|$. This approximation would be very strong specifically when $E^2 \ll 2\Delta$. We know that Fermi-Dirac distribution function always limits the Landauer's formula and therefore the conductance in the range of few $k_B T$ around the Fermi energy which is $25.8 \, meV$ in the room temperature. Meaning that $E$ is confined to few $k_B T$. If we take this explanation into account for $\Delta > 500 \, \mu eV$, the differences between $\left| \sqrt{E^2 - \Delta^2} \right|$ and $\left| \frac{E^2}{2\Delta} - \Delta \right|$ would be less than one percent in the room temperature (which would be even lower in the lower temperatures). Therefore this substitution is perfectly applicable when $\Delta > 500 \, \mu eV$ which is a practical laboratory value. Hence, using this extension, the number of the modes in combination with the Landauer's formula indicates the conductance of the Silicene as:

$$G = -\frac{2q^2}{h}\frac{W}{\pi\hbar v_F}\left(\int\limits_0^{+\infty}dE\left|\frac{E^2}{2\Delta}-\Delta\right|\frac{\partial}{\partial E}\left(\frac{1}{1+e^{\frac{E+E_F}{k_BT}}}\right) - \int\limits_0^{+\infty}dE\left|\frac{E^2}{2\Delta}-\Delta\right|\frac{\partial}{\partial E}\left(\frac{1}{1+e^{\frac{E-E_F}{k_BT}}}\right)\right) \quad (14)$$

Employing partial integration methods and changing variables, the equation (14) could be written as the equation (15) which indicates the general analytical conductance model of the Silisece as a 2D channel in the Silicene based FET devices.

$$G = \frac{Wq^2}{\Delta\pi^2\hbar^2 v_F}(k_BT)^2\left(\Im_1(-\eta)-\Im_1(\eta)\right) \quad (15)$$

Figure 5 shows the variation of the Silicene conductance versus the gate voltages in the different temperatures and zero bias between about the absolute and room temperatures based on the equation (15). The conductance of the Silicene is expected to increase with the gate voltage. The increment rate is higher around the neutrality point and it remains more stable out of the neutrality point (degenerate regime). The minimum conductance increases with the temperature, whereas the temperature does not affect the conductance in the degenerate regime.

Employing similar method, for the single layer Graphene with zero gap band structure and the linear energy dispersion ($E = \hbar v_{FG}|k|$), the conductance of the Graphene could be written as $G_G = \frac{Wq^2}{\pi^2\hbar^2 v_{FG}}(k_BT)(\Im_1(-\eta)+\Im_1(-\eta))$ where $v_{FG} = 1\times 10^6\ ms^{-1}$ [69] for the Graphene. Figure 6 shows the conductance of the Graphene and Silicene (with the same size) plotted based on the presented analytical models in this paper for two different temperatures: T=150K and room temperature. It is clear that the amplitude of the minimum conductance of the Silicene is close to the Graphene which may be explained by the similar band structure in zero gate voltage where the band structure of the Silicene and Graphene are linear. However, in the out of the neutrality point, the conductance of the Graphene is about ten times higher than Silicene.

Graphene and Silicene temperature dependence are different in the out of the neutrality point as shown in figure 6. Although conductance of Silicene remains stable with increasing the temperature, it is reduced for Graphene in the degenerate regime. However, the minimum conductance shows similar trends in the both cases as it is increased by temperature.

The temperature dependence of the conductance and the total carrier of the Silicene in the different gate voltages ($vg = 0.01, 0.1, 0.2, 1\ V$) has been demonstrated in figure 7. The total carriers and the conductance are increased by the temperature for the low gate voltage regime ($vg = 0.01, 0.1\ and\ 0.2\ V$); despite it is remained stable in higher gate voltages ($vg = 1\ V$).

Presented transport models for the Silicene can provide a better understanding toward the Silicene channel behavior in the field-effect transistor configuration which can be employed to model the Silicene based transistors for the integrated circuits.

## V. CONCLUSION

Silicene is a single layer of silicon atoms with $sp^3$ bonds in the honeycomb lattice structure. More compatibility of the Silicene with the current semiconducting technology and some other advantages such as a new and promising alternative for the spintronic and the nanomagnetic materials have attracted considerable scientific attention in very recent years. In addition, despite the single layer Graphene which owns zero band gap in the Dirac point, Silicene owns a tunable band gap from zero to semiconducting region, which make it better choice for the FET technology. However, stability of this material was an issue that the recent experimental studies showed that it could be stable with a small buckling. However, its electronic properties such as the density of states, the total carriers and the conductance as well as their temperature dependence and comparison with the electronic properties of the Graphene have not been investigated analytically yet. In this paper, the analytical transport model of the Silicene, as channel in the field effect transistor configuration was presented. The model shows increment in the total carrier concentration and the conductance with the gate voltage as expected for the conventional semiconductors which affected by the temperature around the neutrality point. The minimum conductance is increased by the temperature whereas it remains stable in the degenerate regime. The Silicene total carrier and conductance are almost affectless by the temperature in the higher gate voltages. Comparing with Graphene conductance, Silicene shows lower conductance in similar size and thr temperature. This model could be employed to model the Silicene FET based integrated circuits.

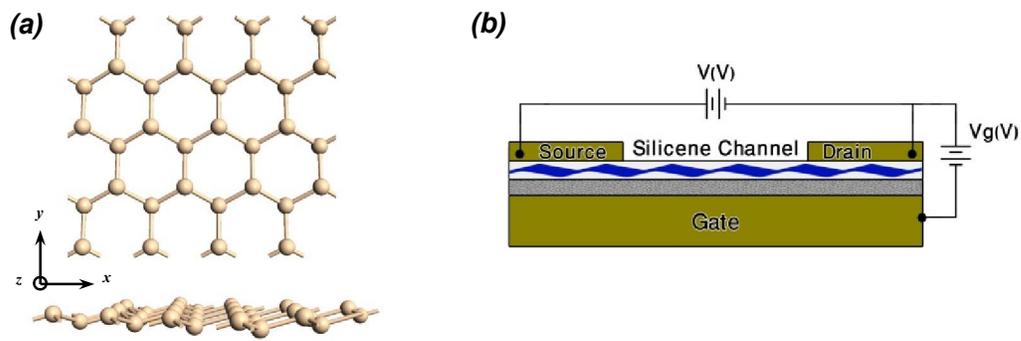

Fig. 1. (a) Silicene molecular structure; (b) Silicene based FET structure

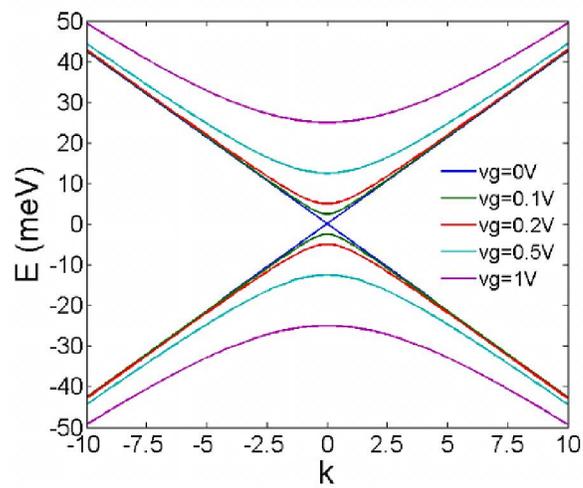

Fig. 2. Silicene band structure in the absence and presence of the perpendicular electrical field

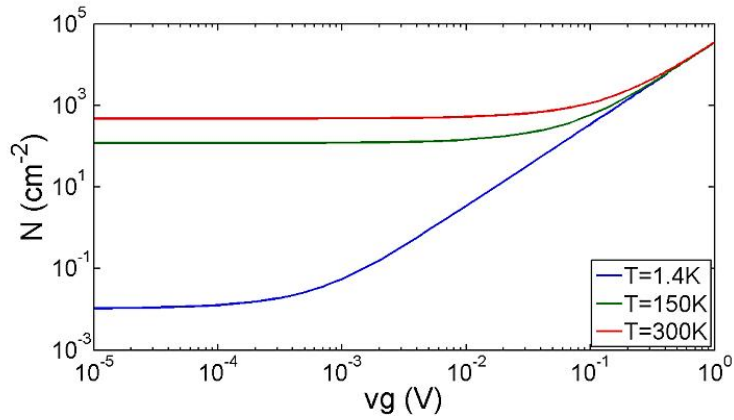

Fig. 3. Total carrier concentration of the Silicene vs. the gate voltage in the different temperatures based on eq (6)

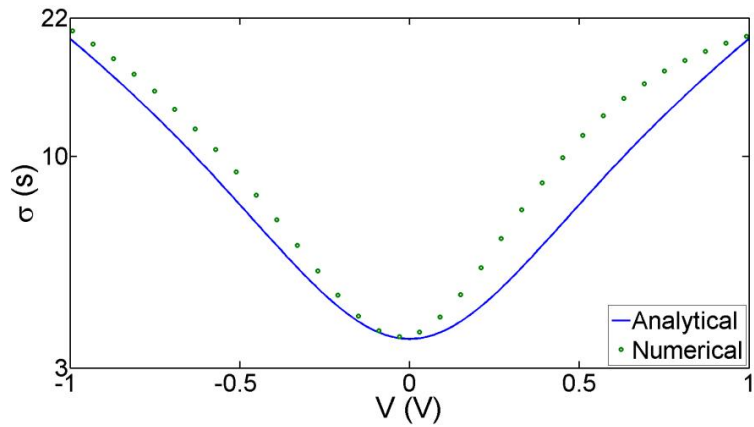

Fig. 4: The analytical (solid line) and the numerical (dots) conductivity of the Silicene in the different bias voltages and zero gate voltage

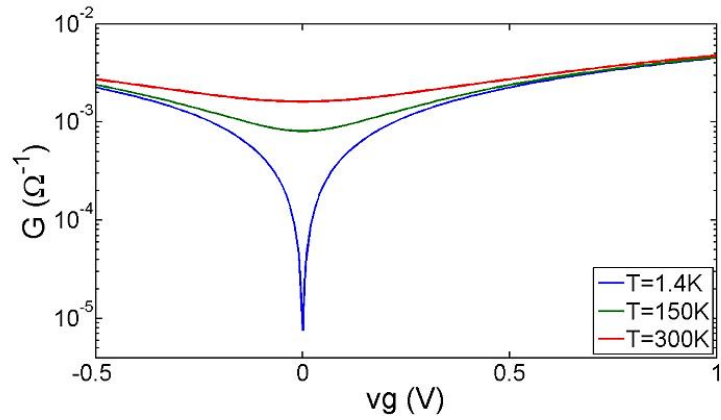

Fig. 5. The conductance of the Silicene vs. the gate voltage in the different temperatures

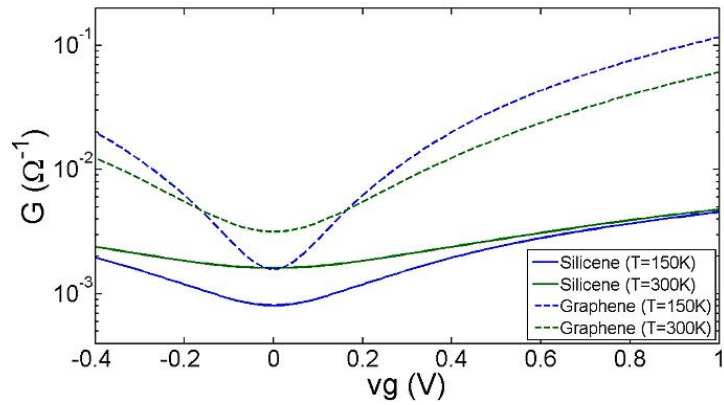

Fig. 6. Comparison of the conductance of the Silicene and single layer Graphene vs. the gate

voltage in T=150K and the room temperature with the similar size

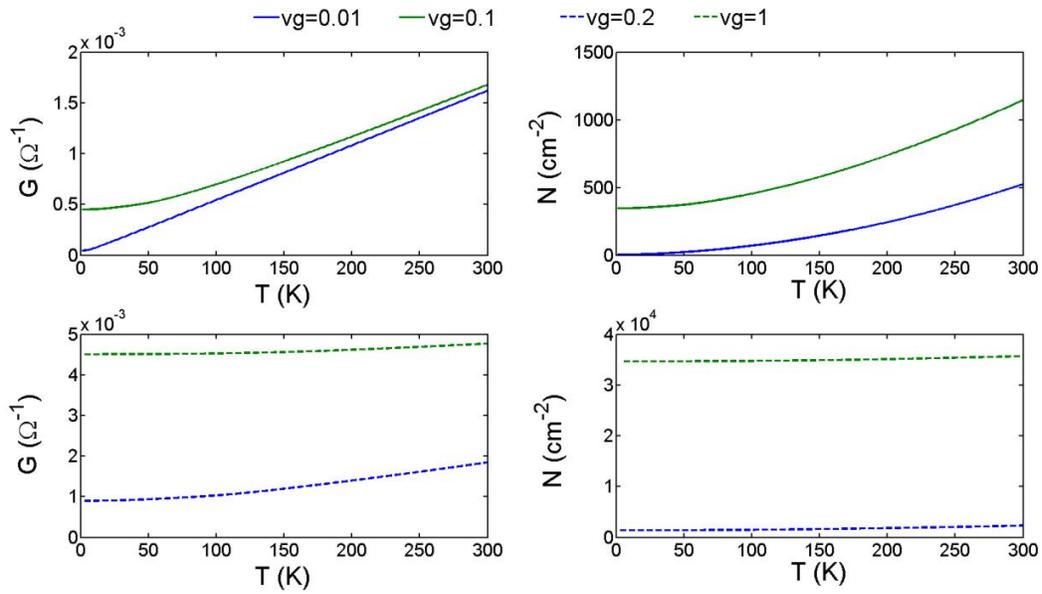

Fig. 7. The conductance (a) and the total carrier (b) of the Silicene vs. the temperatures in the four different gate voltages (0.01, 0.1, 0.2 and 1 V)

## Graphical Abstract

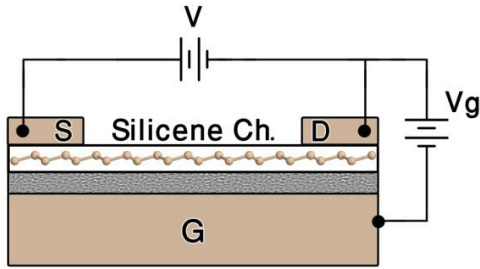 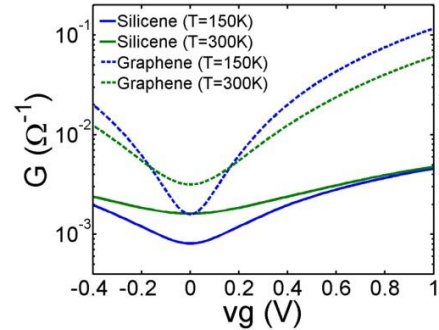

In this paper, the analytical electrical transport model of the Silicene, a single layer of $sp^3$ bonded silicon atoms in the honeycomb lattice structure as a channel in the field effect transistor configuration is presented. Although the carrier concentration of the Silicene shows similar behavior to Graphene, there are some differences in the conductance behavior. Presented analytical model is in good agreement with the numerical conductance calculation based on the implementation of NEGF coupled to the DFT.